\documentclass[12pt]{article}
\usepackage{latexsym}
\usepackage{epsfig}

\usepackage{graphicx}
\usepackage{epstopdf}

\usepackage{epsfig,amssymb,amsmath,amscd}

\newcommand{\bq}{\begin{equation}}
\newcommand{\eq}{\end{equation}}
\newcommand{\bqa}{\begin{eqnarray}}
\newcommand{\eqa}{\end{eqnarray}}
\newcommand{\ben}{\begin{enumerate}}
\newcommand{\een}{\end{enumerate}}
\newcommand{\bc}{\begin{center}}
\newcommand{\ec}{\end{center}}
\newcommand{\bqb}{\begin{eqnarray*}}
\newcommand{\eqb}{\end{eqnarray*}}

\def\pr#1#2#3{Phys. Rev. ${\bf{#1}}$, #2 (#3)}

\def\pl#1#2#3{Phys. Lett. ${\bf{#1}}$, #2 (#3)}

\def\np#1#2#3{Nucl. Phys. ${\bf{#1}}$, #2 (#3)}

\def\jmp#1#2#3{J. Mod. Phys. ${\bf{#1}}$, #2 (#3)}

\begin{document}
\pagenumbering{arabic}
\thispagestyle{empty}
\def\thefootnote{\fnsymbol{footnote}}
\setcounter{footnote}{1}

\vspace*{2cm}
\begin{flushright}
October 13, 2018\\
 \end{flushright}
\vspace*{1cm}

\begin{center}
{\Large {\bf
Higgs boson structure from the shape
of the cross section in exchange processes.}}\\
 \vspace{1cm}
{\large F.M. Renard}\\
\vspace{0.2cm}
Laboratoire Univers et Particules de Montpellier,
UMR 5299\\
Universit\'{e} de Montpellier, Place Eug\`{e}ne Bataillon CC072\\
 F-34095 Montpellier Cedex 5, France.\\
\end{center}

\vspace*{1.cm}
\begin{center}
{\bf Abstract}
\end{center}

We show how the Higgs boson exchange processes may indicate the
occurence of special Higgs boson structures (substructures or
peculiar interactions with dark matter) from a possible modification
of the s-dependence of their cross section. We illustrate the 
simplest example with the $\mu^+\mu^-\to f\bar f$ process.

\vspace{0.5cm}

\def\thefootnote{\arabic{footnote}}
\setcounter{footnote}{0}
\clearpage

\section{Introduction}

The existence of a peculiar structure of the Higgs boson is a
BSM possibility which has been considered for various reasons.
In SM the peculiar fermion spectrum, with enormously different mass scales,
can be taken into account within the Higgs description by using an 
adequate set of couplings with the Higgs doublet, but its origin is not explained.\\
The origin of mass has been discussed since a long time,
see for example \cite{Wilczek, Hansson,Hosek}.\\
Among the various possibilities an extension of the SU(3)*SU(2)*U(1) 
gauge structure may be done in order to
describe the presence of 3 families.
But even if this can differentiate the families the description 
of the specific mass spectrum for each familiy does not seem  quantitatively trivial.\\
The nearby values of the top quark and Higgs boson masses 
can suggest some common origin (for example a substructure)  but 
the very different scales for the masses of the other fermions
would require additional peculiar structures;
\cite{comp, Hcomp2,Hcomp3,Hcomp4,partialcomp}.\\
On another hand the Higgs boson may be a mediator between the standard set and
a new set, for example between our usual world and a new world responsible for
the dark matter (DM); \cite{DMrev,DMmass,DMexch}.\\
Like in the hadronic case with QCD, contributions to the mass
may occur from the various interactions of each fermion,
the SM ones and the ones from a new sector possibly through
the Higgs mediation.
Interferences with different orders of virtual effects
between the SM gauge group and the new structure could create
these different fermionic mass scales.\\

Our point is that, as a consequence of such an origin of mass,  an
energy dependence of the Higgs coupling constants may be generated.
With the usual operators connecting fermion and Higgs 
fields one describes the fermion mass $m_f$ and the $Hff$ coupling
$g_{Hff}$ but the above structures may create off-shell
dependences; an effective mass $m_f(s)$ \cite{trcomp},
like in the QCD case,
and an s-dependent $Hff$ coupling $g_{Hff}(s)$ when the Higgs boson
is off-shell. 
The relation between $g_{Hff}$ and $m_f$ may differ from the SM case
like in \cite{gfmf} but with specific s-dependences.\\
Simple high order virtual corrections (with sufficiently
large couplings and low scales in order to produce visible effects) may occur.
A trivial case is the presence of resonant structures
related to additional Higgs bosons, but richer substructure dependences may
also appear. \\

Effects of H and t compositeness in several processes have already been studied, 
\cite{trcomp,eettZ}.
But the situation may be different for light fermions the smallness of their
masses implying a less direct Higgs connection.\\
An other aspect in this domain concerns the whole SU(2) structure 
of the Higgs states. In SM this concerns the goldstone triplet $G^{0,\pm}$.
If the CSM hypothesis, \cite{CSMrev}, is still valid and the equivalence principle
maintained this should determine the properties of the longitudinal
$W_L$ and $Z_L$ states, \cite{equiv}.\\
So our  point will also be to check if 
the above mentioned new structures affect similarly H and G, and
therefore the $W_L$ and $Z_L$ states and their exchange processes.\\

In the present paper we want to explore how one can test these various assumptions.
Obviously the best way to test these s-dependent  Higgs properties is to study
the Higgs exchange processes.\\
The magnitudes of the corresponding cross sections should therefore be sufficiently large.
With this aim a fermion-antifermion system should be polarized
in order to have only equal helicities (as required by the Higgs
coupling).\\
We will illustrate the case of $\mu^+\mu^-\to f\bar f$
processes assuming that adequate polarized muon beams will be available.
About such possibilities see \cite{mumucoll} and \cite{mumupol}.\\
We will explore the sensitivity to the mentioned new Higgs structures
by introducing an s-dependence through adequate form factors and by looking
at the resulting s-shape of the cross section.\\

Contents: In Section 2 we recall the precise expressions of the SM helicity
amplitudes of the $\mu^+\mu^-\to f\bar f$ process; in Section 3 
we introduce new structures for the $H$ and $G^0$ couplings and we give
illustrations for the $\mu^+\mu^- \to b\bar b$ polarized cross sections.\\

\section{The $\mu^+\mu^-\to f\bar f$ process}

In SM, at Born level, this process involves both gauge and Higgs boson exchanges.
In the unpolarized case the dominant process is the gauge boson ($\gamma,Z$)
exchange; the Higgs ($H$) and Goldstone($G^0$) boson exchanges are reduced by the 
${m_{\mu}\over m_W}$ and ${m_f\over m_W}$ factors. 
If one selects equal helicities separately in the initial and final states
the gauge amplitudes are reduced by the $m_{\mu}\over\sqrt{s}$ 
and $m_f\over\sqrt{s}$ factors and become of comparable size to that of the 
($H, G^0$) exchange.\\
So we will consider the  
$F^{\gamma, Z, H, G^0}(\lambda_{\mu^+}\lambda_{\mu^-}\lambda_f\lambda_{\bar f})$
amplitudes with $\lambda_{\mu^+}=\lambda_{\mu^-}$
and $\lambda_f=\lambda_{\bar f}$.\\
We will illustrate the behaviour of the polarized cross sections

\bq
\sigma={pN_c\over 32\pi ls}\int d\cos \theta \sum |F|^2
\eq
where $p,l$ are the initial and final center of mass momenta; $N_c$ is the
colour number of the fermion $f$.
In $\sum |F|^2$, we will consider either (RR) or (LL) initial states with
$\lambda_{\mu^+}=\lambda_{\mu^-}=\pm{1\over2}$ and sum over both
(RR) and LL) final states with $\lambda_f=\lambda_{\bar f}=\pm{1\over2}$,
i.e.\\
$\sum |F|^2=|F_{RRRR}|^2+|F_{RRLL}|^2$ leading to $\sigma_R$,
or
$|F_{LLLL}|^2+|F_{LLRR}|^2$ leading to $\sigma_L$,
which will lead to the same value after integration over $\cos \theta$.\\
Each amplitude is the sum of 4 terms corresponding to photon, $Z$, $H$ and 
$G^0$ exchanges.\\ 

\bq
F^H_{LLLL}=F^H_{LLRR}=F^H_{RRLL}=F^H_{RRRR}=-4e^2g_{H\mu\mu}g_{Hff}{lp\over D_H(s)}
\eq
\bq
F^G_{LLLL}=-F^G_{LLRR}=-F^G_{RRLL}=F^G_{RRRR}=-4e^2g_{G\mu\mu}g_{Gff}{l^0p^0\over D_G(s)}
\eq
with 
\bq
g_{Hff}={-m_f\over 2s_Wm_W}~~~~g_{Gff}={-im_fI^f_3\over s_Wm_W}
\eq

and for $V=\gamma,Z$ 
\bq
F^V_{LLLL}=F^V_{RRRR}={e^2m_{\mu}m_f\over D_V(s)}[(g^V_{\mu L}-g^V_{\mu R})
(g^V_{f L}-g^V_{f R})-(g^V_{\mu L}+g^V_{\mu R})(g^V_{f L}+g^V_{f R})\cos\theta]
\eq
\bq
F^V_{LLRR}=F^V_{RRLL}={-e^2m_{\mu}m_f\over D_V(s)}[(g^V_{\mu L}-g^V_{\mu R})
(g^V_{f L}-g^V_{f R})+(g^V_{\mu L}+g^V_{\mu R})(g^V_{f L}+g^V_{f R})\cos\theta]
\eq 

with
\bq
g^{\gamma}_{f L}=g^{\gamma}_{f L}=Q_f
\eq
\bq
g^{Z}_{f L}={I^f_3-Q_fs^2_W\over s_Wc_W}~~~~g^{Z}_{f R}={-Q_fs_W\over c_W}
\eq

The illustrations will be done for the bottom quark ($f=b$) with
$\mu^+\mu^-\to \gamma,Z,H,G^0 \to b\bar b$. See Fig.1 (solid line for SM)
showing the s-dependence of the angular integrated cross section
$\sigma_L=\sigma_R$ with the peaks at threshold, at $s=m^2_Z$
and at $s=m^2_H$ due to photon, $Z$, $G^0$ and 
$H$ exchanges.\\

\section{New structures for $H,G^0$ couplings}

We now want to explore the sensitivity of the cross section to a modification of
the Higgs coupling or of both $H,G^0$ couplings.\\
We use a test form for the modification of the product of couplings as suggested
by possible loop (or compositeness) structure.\\

\bq
f(s)=1+clog(1+{s\over m^2_0})
\eq

A first illustration is given in Fig.1(up) by multiplying only the $H$ term
or both $H, G^0$ terms by $f(s)$ with $c=1$ and $m_0=0.1$ TeV. \\
One can see the separate effects of the modification of $H$ contribution
or of a modification of both  $H$ and $G^0$ contributions while the
$\gamma,Z$ ones keep their SM forms.
The chosen $f(s)$ modification in eq.(9) is totally arbitrary. It just shows how
the SM Breit-Wigner shapes can be differently modified, especially when the complete
Higgs sector (with $H$ and $G^0$ contributions) is affected.\\

As a second illustration we impose the SM value to the $H$ couplings at $s=m^2_H$ and 
to the $G^0$ couplings at $s=m^2_Z$, using
 
\bq
f_H(s)=1+clog[{s+m^2_0\over m^2_H+m^2_0}]
\eq   

\bq
f_G(s)=1+clog[{s+m^2_0\over m^2_Z+m^2_0}]
\eq 
with $m^2_0=m^2_b$.  

In Fig.1(down) we can also compare the basic SM shape to that of the two cases where $f_H(s)$
or both  $f_H(s)$ and $f_G(s)$ are applied, leading to notably different modifications. \\

From these simple examples one can expect that precise measurements of the shape of
the s-dependence of the $\mu^+\mu^- \to b\bar b$ cross section could reveal what type
of Higgs structure may occur.\\
   
More complex s-dependences with resonances, threshold effects or other dynamical
features,... may also appear.\\

\section{Conclusion}

In this study we wanted to test the sensitivity of Higgs exchange processes to
the occurence of an Higgs boson structure (substructure, DM cloud,....) even if the on-shell coupling
constants agree with the SM prediction (no change in the decay branching
ratios, neither in the on-shell production rates).\\
As a simple example we have considered the $\mu^+\mu^-\to \gamma,Z,H,G^0 \to b\bar b$
process. In order to increase the sensitivity to the $H,G^0$ contributions one
should use polarized $\mu^{\pm}$ beams with $\lambda_{\mu^+}=\lambda_{\mu^-}$
and restrict the final states to $\lambda_b=\lambda_{\bar b}$.\\
We have illustrated how modifications (form factors) of the $Hff$ couplings and possibly
of both $Hff$ and $G^0ff$ ones, 
even preserving their SM on-shell value, can notably affect the shape of the s-dependence of the
corresponding polarized cross section.\\
In order to be applicable to precise analyses this study should be completed by taking into
account high order SM effects (bremsstrahlung, radiative corections)
and experimental detection features.\\

In addition to this idealistic example other possibilities may be considered. 
One possibility may be multibody production processes
for example $t\bar t f\bar f$ in $e^+e^-$ or hadronic collisions. 
The contribution of the $H,G^0$ exchanges (with $t\bar t H,G^0$ production followed by
$H,G^0\to f\bar f$) would also be favored by selecting equal helicities for both
$t\bar t$ and $f\bar f$ systems.
Such a study would require a very complex experimental analysis of the
$t\bar t f\bar f$ final state. See for example \cite{ttbb} for LHC
and \cite{ee} for future $e^+e^-$ colliders. \\

\newpage

\begin{figure}[p]
\vspace{-4cm}
\[
\hspace{-2cm}\epsfig{file=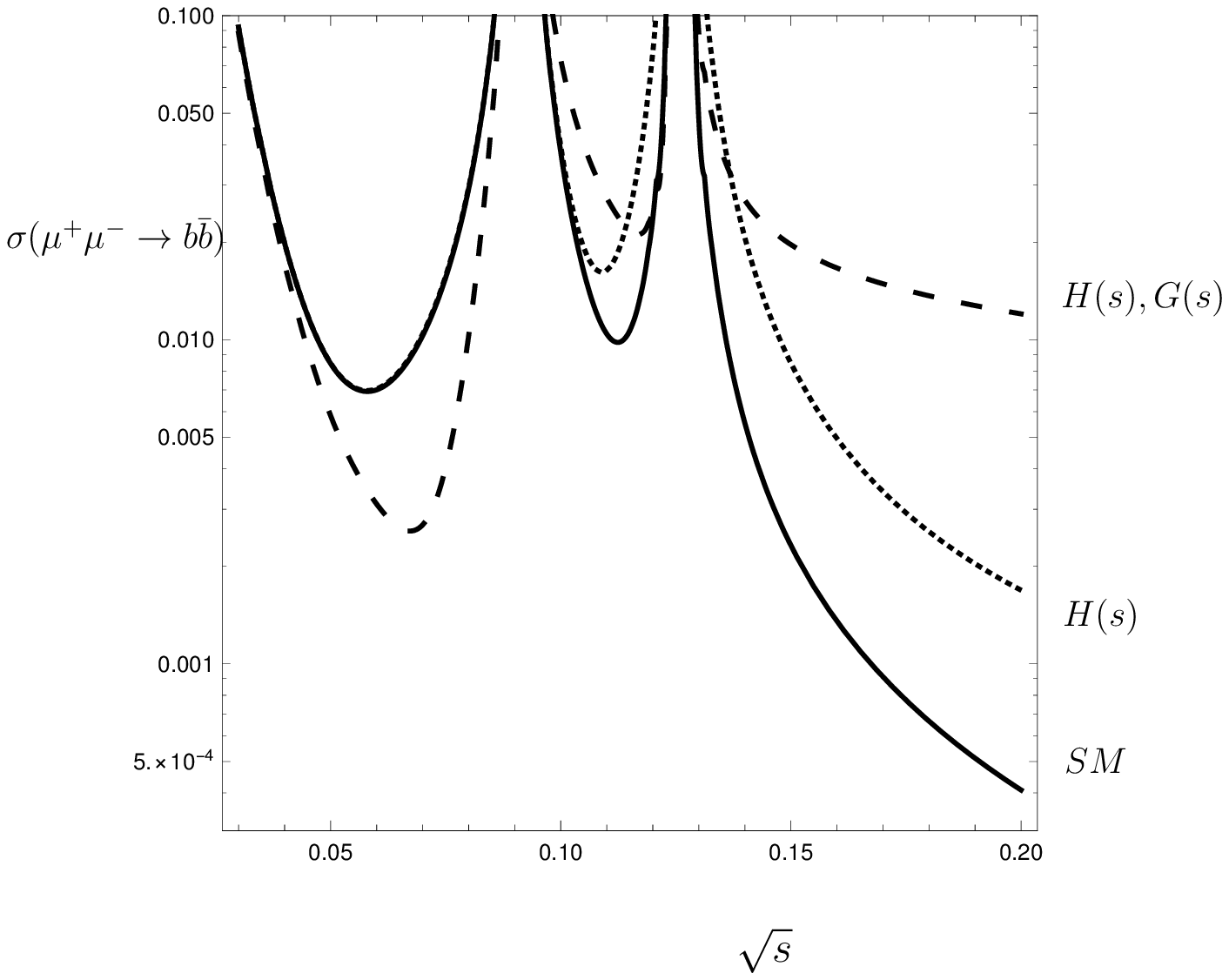 , height=20.cm}
\]\\
\vspace{-13cm}

\[
\hspace{-2cm}\epsfig{file=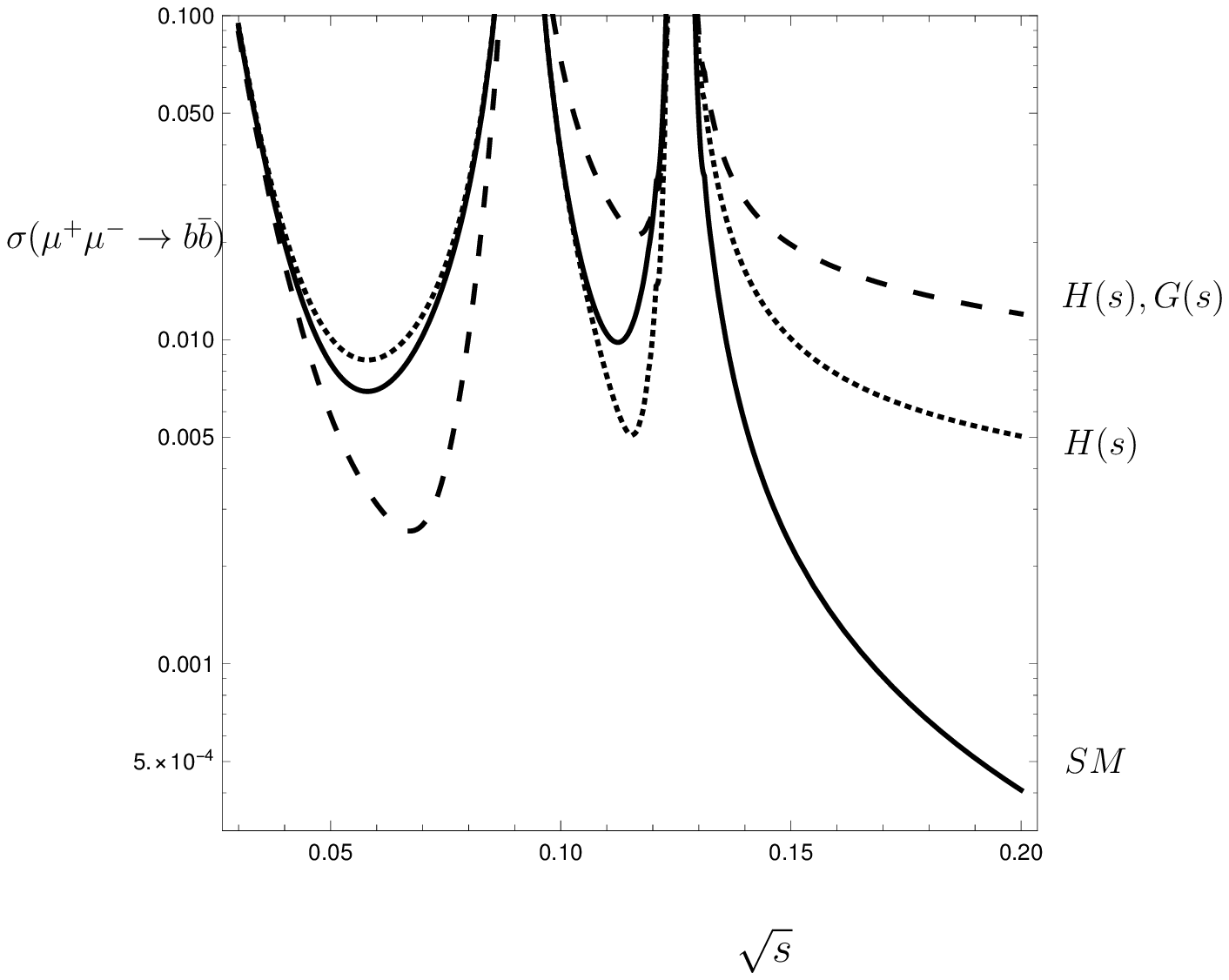 , height=20.cm}
\]\\
\vspace{-10cm}
\caption[1] {Shapes of $\sigma(\mu^+\mu^-\to b\bar b)$ without normalization at $s=m^2_H$ 
and at $s=m^2_Z$ (up);
with normalization at $s=m^2_H$ and at $s=m^2_Z$ (down).}

\end{figure}


\end{document}